\documentclass{article}
\usepackage{epsf}
\usepackage{amsmath,amssymb}
\usepackage{graphicx}
\def\abstract#1{\vskip 7mm 
        \begin{center}{\large Abstract}\par \smallskip
                \begin{minipage}[c]{12cm}
                        \small #1
                \end{minipage}
        \end{center}
}
\def\title#1{\begin{center}{\Large\bf #1}\end{center}}
\def\author#1{\vskip 5mm \begin{center}{#1}\end{center}}
\def\address#1{\begin{center}{\it #1}\end{center}}
\makeatletter

\@ifundefined{lesssim}{\def\lesssim{\mathrel{\mathpalette\vereq<}}}{}
\@ifundefined{gtrsim}{}{}
\def\vereq#1#2{\lower3pt\vbox{\baselineskip1.5pt \lineskip1.5pt
\ialign{$\m@th#1\hfill##\hfil$\crcr#2\crcr\sim\crcr}}}
\makeatother

\begin{document}

\hfill TU-638\\
\title{%
  Cosmic Microwave Background Anisotropy\\ 
 in models with Quintessence\footnote{
   To appear in the proceedings of "Frontier of Cosmology and
   Gravitation" (YITP, Kyoto, April 25-27, 2001).  This presentation
   is based on the work with M.\ Kawasaki and T.\ Moroi
   \cite{kawasaki_moroi_takahashi_2001a,kawasaki_moroi_takahashi_2001b}.}
 }
\author{%
Tomo Takahashi
}
\address{
  Department of Physics,  Tohoku University\\
  Sendai 980-8578, Japan
  }
\abstract{
  We study the Cosmic Microwave Background (CMB) anisotropies
  produced by cosine-type quintessence models.  In our analysis,
  effects of the adiabatic and isocurvature fluctuations are both
  taken into account. For purely adiabatic fluctuations with scale
  invariant spectrum, we obtain a stringent constraint on the model
  parameters using the CMB data from COBE, BOOMERanG and MAXIMA.
  Furthermore, it is shown that isocurvature fluctuations have
  significant effects on the CMB angular power spectrum at low
  multipoles in some parameter space, which may be detectable in
  future satellite experiments.  Such a signal may be used to test
  the cosine-type quintessence models.
}

\section{Introduction}
Recent cosmological observations suggest that there exists a dark
energy which must be added to the matter density in order to reach the
critical density. Although the cosmological constant is usually
assumed as the dark energy, in the past years, a slowly evolving
scalar field, dubbed as ``quintessence'' has been proposed as the dark
energy \cite{caldwell_et_al_1998} and has been been studied by many
authors \cite{quintessence}. There are some differences between the
cosmological constant and quintessence. Firstly, for the quintessence,
the equation-of-state parameter $\omega_Q = \rho_Q / p_Q $ varies with
time, whilst for the cosmological constant, it remains a fixed value
$\omega_{\Lambda} = - 1$.  Secondly, since the quintessence is a
scalar field, it fluctuates.

One of the observational effects produced by the existence of the
quintessence is the CMB anisotropies.  In many cases, the quintessence
dominates the universe at late times ($z \sim O(1)$, with $z$ being
redshift) after the recombination.  At that epoch the gravitational
potential is changed because the equation of state for quintessence is
different from that for non-relativistic matter, which leads to an
enhancement of the CMB anisotropies at large angular scales $l
\lesssim 10$ due to the late-time integrated Sachs-Wolfe effect.  The
quintessence also changes the locations of the acoustic peaks in the
CMB angular power spectrum because the projection of the horizon at
last scattering onto the present sky is enlarged compared with models
with the cosmological constant.  Furthermore, the initial fluctuations
of the quintessence fields are generated during inflation. These
fluctuations behave as isocurvature mode \cite{Abramo} and hence the
quintessence may have both adiabatic and isocurvature perturbations.

\vspace{-2mm}
\section{CMB anisotropy in models with cosine-type quintessence}
The quintessence model we take here is the cosine-type quintessence
which has the potential
\vspace{-2mm}
\begin{eqnarray}
     V(Q) = \Lambda^4
     \left[ 1 -
         \cos\left(\frac{Q}{f_Q}\right) \right]
     = 2\Lambda^4
     \sin^2\left(\frac{Q}{2f_Q}\right), 
     \label{V_q} 
\end{eqnarray}  
where $f_Q$ and $\Lambda$ are model parameters. This type of potential
can be generated if the quintessence field is a pseudo Nambu-Goldstone
boson. In this class of models, effective mass of the quintessence
field is always of $O(\Lambda^2/f_Q)$. 

Since the quintessence is a scalar field, its amplitude may have
position-dependent fluctuations.  To investigate its behavior, we
decompose the $Q$ field as $Q (t, \vec{x}) = \bar{Q} (t) + q (t,
\vec{x})$, where $q$ is the perturbation of the amplitude of the
quintessence field. Before we consider the fluctuations of the
quintessence fields, we discuss the dynamics of the zero mode of the
quintessence fields briefly. For details, see
Ref.\cite{kawasaki_moroi_takahashi_2001a}.

The zero mode $\bar{Q}$ obeys the equation of motion $ \ddot{\bar{Q}}
+ 3 H \dot{Q} + dV/dQ=0$ where the dot represents the derivative with
respect to time $t$. When the relation $H\lesssim\Lambda^2/f_Q$ is
satisfied, the quintessence field starts to oscillate.  If the
combination $\Lambda^2/f_Q$ is large, the oscillation starts earlier
epoch and the quintessence field undergoes many oscillations until the
present time.  We call the parameter space where the oscillation of
the quintessence is significant as ``oscillatory region.''  When the
quintessence field starts to oscillate earlier, it dominates the
energy density of the universe from earlier epoch.  This has
significant implications to the CMB power spectrum.

Now we discuss the the fluctuations of the quintessence fields.  There
are two origins of non-vanishing $q$ at the present time; one is
non-vanishing gravitational potential $\Psi$ and the other is
primordial $q$ itself.  We call these modes as ``adiabatic'' and
``isocurvature'' modes, respectively.  In the actual situations, both
modes may exist.  Since these modes are uncorrelated, the total
fluctuation can be decomposed into two modes which originate to
primordial fluctuations in $\Psi$ and $q$, respectively:
\vspace{-1mm}\\ 
\begin{minipage}{7cm}
\begin{eqnarray}
     {\rm Adiabatic} :
    \left\{
        \begin{array}{l}
            \Psi (a\rightarrow 0) \neq 0  \notag \\
              q (a\rightarrow 0) = 0
        \end{array}
    \right. ,
\end{eqnarray}
\end{minipage}
\hfill
\begin{minipage}{7cm}
\begin{eqnarray}
    {\rm Isocurvature} :
    \left\{
        \begin{array}{l}
            \Psi (a\rightarrow 0) = 0 \\
            q (a\rightarrow 0) \neq 0
        \end{array}
    \right. .
    \label{iso(t=0)}
\end{eqnarray}
\end{minipage}
\vspace{2mm}\\ We assume that there is no entropy fluctuations between
any of two energy components other than that of quintessence.  Thus,
the adiabatic mode given with the initial condition (\ref{iso(t=0)})
corresponds to the conventional adiabatic initial condition.  On the
contrary, the mode with the initial condition (\ref{iso(t=0)}) is
called ``isocurvature mode,'' since the total density fluctuation and
the potential $\Psi$ vanishes as $a\rightarrow 0$ if this is the only
source of the fluctuation in the early universe.  To parameterize the
relative size of adiabatic and isocurvature contributions, it is
convenient to define $r_q$, the ratio of the primordial value of
$\tilde{q}$ to that of the gauge-invariant variable $\tilde{\Psi}$ at
the radiation-dominated universe \vspace{-1.7mm}
\begin{eqnarray}
     r_q \equiv
     \frac{\tilde{q}}{M_*\tilde{\Psi}},
     \label{q/Psi}
\end{eqnarray}
where $M_*$ is the reduced Planck scale.  Since the ratio $r_q$
generically depends on the model of the inflation, we treat $r_q$ as a
free parameter in our analysis. (For example, for the chaotic
inflation with $V_{\rm inf}\propto\chi^p$ with $p$ being an integer
$p=2$ $-$ 10, $r_q\simeq 0.3$ $-$ 0.6.
\footnote{
In our analysis, we only consider the case with scale-independent
primordial fluctuation, and $r_q$ is treated as a scale-invariant
quantity.  When $r_q$ has a scale dependence via $\tilde{\Psi}$ and/or
$\tilde{q}$, $r_q(k)$ for the present horizon scale becomes the most
important since the quadrupole anisotropy is most strongly affected by
the isocurvature mode as will be shown below.}
)

\vspace{-2mm}
\subsection{Adiabatic Fluctuation}
Here, we discuss the CMB anisotropy in models with quintessence. To
calculate the CMB angular power spectrum, we used the modified version
of CMBFAST \cite{cmbfast}.  In this section, we consider the adiabatic
modes, so we take $r_q=0$.  

\begin{figure}[t]
  \vspace{-10mm}
  \begin{minipage}{8.5cm}
    \hspace{-10mm}
   \centerline{
    \scalebox{0.7}{
        \includegraphics{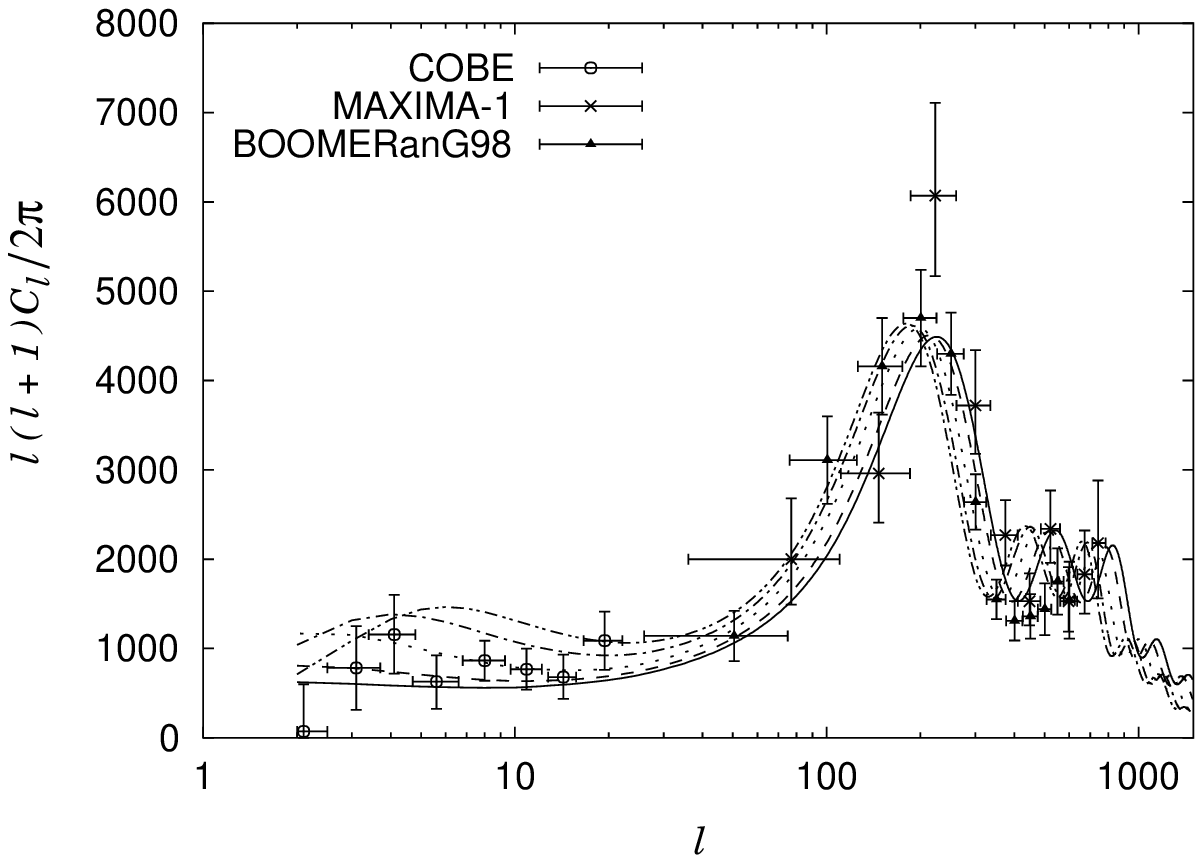}}
}
\vspace{-3mm}
    \caption{The CMB angular power spectrum 
      in models with cosine-type quintessence. The parameters we take
      here, $f_{Q}= 1.8 \times 10^{18}$ GeV, $\Lambda = 2.4 \times
      10^{-12}$ GeV (dashed line), $\Lambda = 3.0 \times 10^{-12}$ GeV
      (dotted line), $\Lambda = 4.0 \times 10^{-12}$ GeV (dash-dot
      line), $\Lambda = 5.0 \times 10^{-12}$ GeV (dash-dot-dot
      line). For comparison, we also show the $\Lambda$CDM case (solid
      line). The cosmological parameters are taken to be $h=0.65$,
      $\Omega_{\Lambda ,Q}=0.7$, $\Omega_{\rm m}=0.3$, $\Omega_{\rm
        b}h^{2}=0.019$, and the initial spectral index is $n=1$.  We
      also show the data points from COBE \cite{cobe},
      BOOMERanG \cite{boomerang}, and MAXIMA \cite{maxima}.}
    \label{fig:Cl(adi)}
  \end{minipage} \hspace{5mm}
  \begin{minipage}{7.5cm}
    \vspace{-15mm}
    \centerline{\scalebox{0.95}{
        \includegraphics{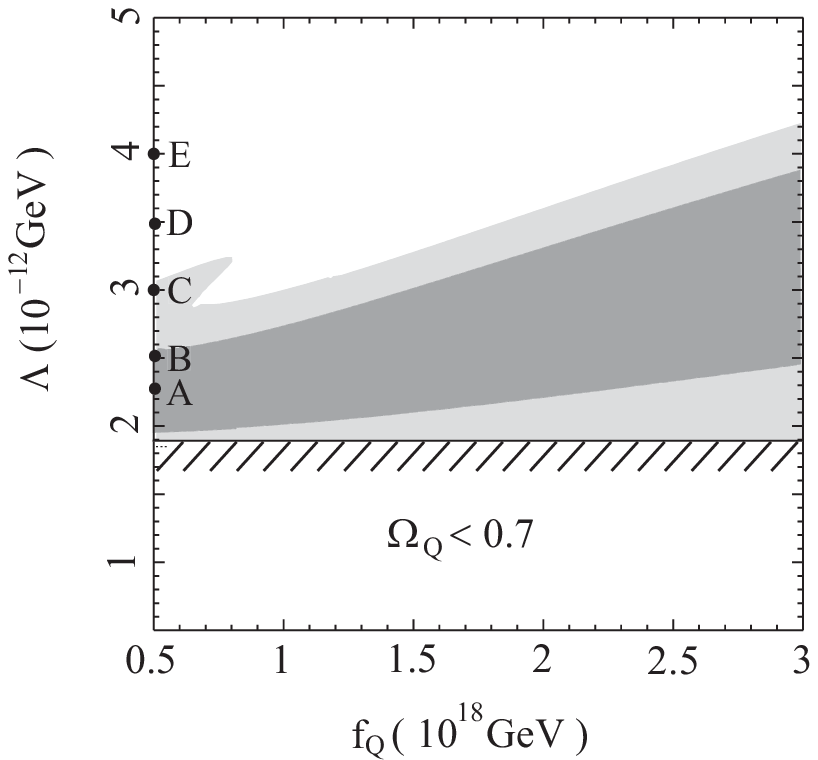}}}
    \caption{Constraints on the parameters $\Lambda$ and $f_{Q}$.
      The lightly shaded regions correspond to 99 \% C.L., and the
      darkly shaded region corresponds to 95 \% C.L. The cosmological
      parameters are the same as in Fig.\ref{fig:Cl(adi)}. The Points
      A $-$ E will be used as representative points in the later
      discussion.}
    \label{fig:chi2}
  \end{minipage}
\end{figure}

Some of interesting features in the CMB
angular power spectrum are discussed in order.  First, let us consider
the locations of the acoustic peaks.  The locations of the peaks
depend on two quantities, the sound horizon at last scattering and the
angular diameter distance to the last scattering surface.
Approximately, the location of $n$-th peak in the $l$ space is
estimated as $l_{n} \simeq (r_\theta (\tau_*)/r_{s}(\tau_{*})) n \pi$
\cite{HuSug}, where $r_{\theta}(\tau_{*}) $ is the angular diameter
distance to the last scattering surface and $r_{s}(\tau_{*})$ is the
sound horizon at last scattering. Since the behavior of the
cosine-type quintessence is almost the same as that of the
cosmological constant until very recently, $r_{s}(\tau_{*})$ is the
same in both cases. However, $r_{\theta}(\tau_{*}) $ in the
quintessence model is different from that in the $\Lambda$CDM models.
Since the quintessence models provides larger total energy density of
the universe than the $\Lambda$CDM models in the earlier epoch, the
angular diameter distance in the quintessence model becomes smaller
than that in the $\Lambda$CDM models.  As we can see from Fig.\ 
\ref{fig:Cl(adi)}, the location of the peaks is shifted to lower
multipole $l$ for the quintessence models.  If we take a parameter in
the oscillatory region, this feature becomes more prominent.  

Next let us consider the height of the acoustic peaks.  Since the
energy density of the quintessence becomes dominant when $z\sim O(1)$,
the late time integrated Sachs-Wolfe (ISW) effect enhances low
multipoles.  Such an enhancement may be more effective in the
quintessence models than in the $\Lambda$CDM models since, in the
quintessence case, the ``dark energy'' (i.e., the quintessence) may
dominate the universe earlier than in the $\Lambda$CDM case.  As a
result, the ratio of the height of the first peak to $C_{10}$ becomes
smaller.  On the contrary, since the quintessence becomes the dominant
component of the universe only at later epoch, pattern of the acoustic
oscillation before the recombination does not change compared to
$\Lambda$CDM models.  Therefore, ratios of the height of the first
peak to those of higher peaks are the same as $\Lambda$CDM models.

Now we discuss constraints on the cosine-type quintessence models from
the observations of COBE, BOOMERanG and MAXIMA.  Constraints on the
parameter $f_Q$ and $\Lambda$ are shown in Fig.\ \ref{fig:chi2}.
When we take a parameter in the oscillatory regions, the quintessence
field becomes dominant component of the universe at earlier epoch.
Namely the late time ISW effect becomes large and the angular diameter
distance to the last scattering surface becomes smaller as mentioned
before.  Therefore, if $\Lambda$ is significantly large, the late time
ISW effect enhances angular power spectrum at low multipoles.  As a
result, the heights of the acoustic peaks are suppressed relative to
$C_l$ with small $l$.

\vspace{-2mm}
\subsection{Isocurvature Fluctuation}
Next, we consider the CMB anisotropy in the case with the isocurvature
mode.  We calculate the CMB anisotropy for various cases, and in Table
\ref{table:c2/c10}, we show the quadrupole $C_2$ normalized by
$C_{10}$.

\begin{figure}
    \begin{minipage}{6cm}
      \vspace{-20mm}
    \centerline{
      \includegraphics[width=7.5cm]{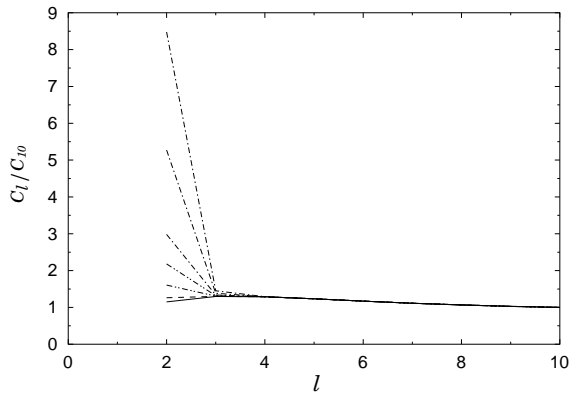}}
\vspace{-3mm}
\caption{$C_l/C_{10}$ for $r_q=0$, 0.5, 1, 1.5, 2, 3 and 4 from
    below with $f_Q=5\times 10^{17}\ {\rm GeV}$ and $\Lambda=3.0\times
    10^{-12}$ GeV.} 
  \label{fig:iso}  
\end{minipage} \vspace{-58mm}  
\end{figure}

If we limit ourselves to the parameter region which is consistent with
the COBE, BOOMERanG, and MAXIMA observations with simple
scale-invariant primordial fluctuation, effect of the isocurvature
mode is quite small as far as $r_q\lesssim 1$.  This is because, in
such cases, there is a severe upper bound on $\Lambda$ to suppress the
late time ISW effect which enhances $C_l$ with small $l$.  As a
result, the quintessence field cannot dominate the universe when $z\gg
1$.  Then, the isocurvature fluctuation in the quintessence density
also becomes a minor effect until very recently. As one can see in
Table \ref{table:c2/c10}, for the best-fit value of $\Lambda$ (i.e.,
for the Point A given in Fig.\ \ref{fig:chi2}), the enhancement of
$C_2$ is about 2 \% even for $r_Q=2$.  If we consider larger value of
$\Lambda$, effect on $C_2$ is more enhanced.  For
$(f_Q,\Lambda)=(5\times 10^{17}\ {\rm GeV},3.0\times 10^{-12}\ {\rm
GeV})$ (i.e., for the Point C given in Fig.\ \ref{fig:iso}, which is
allowed at 99 \% C.L.), we calculate the CMB angular power spectrum
and the result is given in Fig.\ \ref{fig:iso}.  In this case,
$C_2$ can be enhanced by the factor 2.6 if $r_Q=2$.

One should note that angular power spectrum of the CMB anisotropy
strongly depends on the primordial spectrum of the fluctuations.
Thus, the constraint on the $f_Q$ vs.\ $\Lambda$ plane is sensitive to
the scale-dependence of the primordial adiabatic fluctuation which is
determined by the model of the inflation.  Therefore, if we adopt a
possibility of a non-trivial scale dependence of the primordial
fluctuation, the constraint on the $f_Q$ vs.\ $\Lambda$ plane given in
the previous section may be relaxed or modified. If this is the case,
larger value of $\Lambda$ may be allowed and the energy density of the
quintessence field may become significant at earlier stage of the
universe.  
\begin{table}
  \hspace*{\fill}
      \begin{minipage}{8cm}     
        \hspace{0mm}
        \begin{tabular}{lccccc}
          \hline\hline
            {} & {$r_q=0$} & {$r_q=0.5$} & {$r_q=1$} 
            & {$r_q=1.5$} & {$r_q=2$} \\
            \hline
            {A} & {1.31} & {1.31} & {1.32} & {1.33} & {1.34} \\
            {B} & {1.45} & {1.46} & {1.48} & {1.51} & {1.56}  \\
            {C} & {1.15} & {1.26} & {1.61} & {2.18} & {2.98} \\
            {D} & {0.84} & {1.33} & {2.80} & {5.25} & {8.69} \\
            {E} & {0.62} & {1.93} & {5.86} & {12.41} & {21.58} \\
            \hline\hline
        \end{tabular}
        \caption{$C_2/C_{10}$ for several values of 
        $r_q$.  We take $f_Q=5\times 10^{17}\ {\rm GeV}$, and (A)
        $\Lambda=2.3\times 10^{-12}$ GeV, (B) $\Lambda=2.5\times
        10^{-12}$ GeV, (C) $\Lambda=3.0\times 10^{-12}$ GeV, (D)
        $\Lambda=3.5\times 10^{-12}$ GeV, and (E) $\Lambda=4.0\times
        10^{-12}$ GeV.  The Points A $-$ E are indicated
        in Fig.\ \ref{fig:chi2}.}
        \label{table:c2/c10}
        \end{minipage}
\end{table}

\vspace{-2mm}
\section{Conclusions and Discussion}
We have studied the CMB anisotropies produced by cosine-type
quintessence models.  In particular, effects of the adiabatic and
isocurvature fluctuations have been both discussed.

For purely adiabatic fluctuations with scale invariant spectrum, the
existence of the quintessence suppresses the relative height of the
first acoustic peak of the angular power spectrum compared with the
$\Lambda$CDM case.  This is because, in the quintessence models, the
``dark energy'' due to the quintessence may dominate the universe
earlier than the cosmological constant case, and hence the late time
ISW effect becomes more effective.  As a result, the CMB anisotropy
for large angular scale is more enhanced, which relatively suppresses
the height of the acoustic peaks.  Because of this effect, the CMB
data from COBE, BOOMERanG and MAXIMA have imposed the stringent
constraint on the model parameters of the quintessence models.  We
have also seen that the location of the the first acoustic peak shifts
to lower multipole $l$ compared with $\Lambda$CDM models.

In the case of the isocurvature fluctuations, we have shown that the
isocurvature fluctuations have significant effects on the CMB angular
power spectrum at low multipoles in some parameter space, which may be
detectable in future satellite experiments.  Such effects is also seen
in the tracker-type models \cite{kawasaki_moroi_takahashi_2001b}.  This
signal may be used to test the quitessence model, combining with the
global shape of the CMB angular power spectrum.

\end{document}